\documentclass[aps,prd,twocolumn,floats,nofootinbib,longbibliography,showkeys]{revtex4-1}
\usepackage[dvips]{graphicx}

\usepackage{amsmath}
\usepackage{braket}
\usepackage{amsfonts}
\usepackage[colorlinks=true]{hyperref}
\usepackage{hyperref}
\usepackage{xcolor}

\usepackage{booktabs}

\usepackage{aas_macros}

\usepackage{caption}
\usepackage{graphicx}
\usepackage{subcaption}

\usepackage{bm}
\usepackage{slashed}

\usepackage{changes}

\def\aap{\textit{Astronomy and Astrophysics}}

\def\be{\begin{equation}}
\def\ee{\end{equation}}
\def\bea{\begin{eqnarray}}
\def\eea{\end{eqnarray}}

\def\L{\mathcal{L}}
\def\Lm{\mathcal{L}_m}

\def\t{\tilde}
\newcommand\D{\mathcal D}

\definechangesauthor[name={Olivier}, color=magenta]{OM}
\definechangesauthor[name={Aurelien}, color=cyan]{AH}

\begin{document}
\title{Deriving Entangled Relativity}

\author{Olivier Minazzoli}
\email[]{ominazzoli@gmail.com}
\affiliation{Artemis, Universit\'e C\^ote d'Azur, CNRS, Observatoire C\^ote d'Azur, BP4229, 06304, Nice Cedex 4, France,\\Bureau des Affaires Spatiales, 2 rue du Gabian, 98000  Monaco.}
\author{Maxime Wavasseur}
\affiliation{Departament de Física Quántica i Astrofísica (FQA), Universitat de Barcelona (UB), Carrer de Martí i Franquès, 1, Barcelona, 08028, Spain,\\Artemis, Universit\'e C\^ote d'Azur, CNRS, Observatoire C\^ote d'Azur, BP4229, 06304, Nice Cedex 4, France.}
\author{Thomas Chehab}
\affiliation{Artemis, Universit\'e C\^ote d'Azur, CNRS, Observatoire C\^ote d'Azur, BP4229, 06304, Nice Cedex 4, France.}
\begin{abstract}
Entangled Relativity is a non-linear reformulation of Einstein's theory that cannot be defined in the absence of matter fields. It recovers General Relativity without a cosmological constant in the weak matter density limit or whenever $\Lm = T$ on-shell, and it is also more parsimonious in terms of fundamental constants and units. In this paper, we show that Entangled Relativity can be derived from a general $f(R,\Lm)$ theory by imposing a single requirement: the theory must admit all solutions of General Relativity without a cosmological constant whenever $\Lm = T  \neq 0$ on-shell, though not necessarily only those solutions. An important consequence is that all vacuum solutions of General Relativity without a cosmological constant are limits of solutions of Entangled Relativity when the matter fields tend to zero. 
In addition, we introduce a broader class of theories featuring an \textit{intrinsic decoupling}, which, however, do not generally admit the solutions of General Relativity.
\end{abstract}
\maketitle

\section{Introduction}

Entangled Relativity can be viewed as a non-linear reformulation of General Relativity that fundamentally relies on the presence of matter for its very definition. This framework aligns with Einstein’s original viewpoint that a truly satisfactory theory of relativity should not permit the existence of vacuum spacetimes \citep{einstein:1917co,einstein:1918an,einstein:1918sp,einstein:1921bk,pais:1982bk,hoefer:1995cf}. In fact, the existence of vacuum solutions would imply that inertia—determined by the metric tensor in relativistic theories—could be meaningfully defined even in the total absence of matter. Such a possibility would, in essence, contradict the principle of relativity of inertia \citep{einstein:1917co,einstein:1918an,einstein:1918sp,einstein:1921bk,pais:1982bk,hoefer:1995cf,minazzoli:2024pn}, which Einstein named Mach's principle in \citep{einstein:1918an}. Entangled Relativity is defined from the following path integral \cite{minazzoli:2022ar}
\be
Z = \int [\D g]  \prod_j [\D f_j] \exp \left[-\frac{i}{2 \epsilon^2} \int d^4_g x \frac{\L^2_m(f,g)}{R(g)} \right], \label{eq:ERPI}
\ee
where $\int [\D]$ represents the summation over all distinct field configurations, $R$ denotes the conventional Ricci scalar built from the metric tensor $g$, and $ \mathrm{d}^4_g x := \sqrt{-|g|}  \mathrm{d}^4 x$ corresponds to the spacetime volume element, with $|g|$ being the determinant of $g$. The function $\L_m$ is the Lagrangian density associated with matter fields $f$, which could correspond to the present \textit{standard model of particle physics} Lagrangian density or, more likely, an extension of it. Similar to General Relativity, Entangled Relativity does not impose constraints on the specific matter content of the universe. It also involves the metric tensor, primarily through the conventional \textit{comma-goes-to-semicolon rule} \citep{MTW}, though, in a strict sense, this condition is necessary only in the regime where Entangled Relativity reduces to General Relativity.  

From Eq. (\ref{eq:ERPI}), it is evident that the theory cannot even be defined in the absence of matter, thereby aligning more closely with the principle of relativity of inertia than General Relativity \cite{minazzoli:2024pn}.

Entangled Relativity was initially introduced in \cite{ludwig:2015pl} and was formally named in \cite{arruga:2021pr}. The terminology does not refer to the notion of \textit{quantum entanglement}, but stems from the fundamental feature that, within this framework, matter and curvature cannot be treated independently at the level of the action, meaning they are intrinsically \textit{entangled} in the literal sense of the word.

Entangled Relativity has been demonstrated to recover General Relativity as a limiting case in a broad range of classical situations \citep{ludwig:2015pl,minazzoli:2018pr,arruga:2021pr,arruga:2021ep,minazzoli:2021ej,minazzoli:2021cq,minazzoli:2024mo,wavasseur:2025gg,minazzoli:2025ep}. 
Notably, it has been shown to be indistinguishable from, or very close to, the phenomenology of General Relativity in the solar system \cite{ludwig:2015pl,minazzoli:2013pr}, in black hole physics \cite{minazzoli:2021ej,wavasseur:2025gg,minazzoli:2025ep}, in neutron star physics \cite{arruga:2021pr,arruga:2021ep,chehab:2025hl}, and has also been argued to be able to approximate General Relativity with a cosmological constant in cosmology under specific assumptions \cite{minazzoli:2021cq}---although much work remains to be done. This is due to a specific decoupling of the additional gravitational scalar degree of freedom with respect to General Relativity whenever $\Lm=T$ on-shell. This kind of decoupling is known as an \textit{intrinsic decoupling}. It has been originally found in Scalar-Tensor theories \cite{minazzoli:2013pr,minazzoli:2014pr,minazzoli:2014pl}, and is also at play in Entangled Relativity \cite{ludwig:2015pl,minazzoli:2018pr,minazzoli:2021cq,minazzoli:2024mo,arruga:2021pr,arruga:2021ep,minazzoli:2021ej,minazzoli:2022ar,minazzoli:2025ep,wavasseur:2025gg,minazzoli:2025ep}. This suggests that the theory remains a viable alternative, at least in light of current observational constraints, and merits further investigation.

The discovery of the action for the theory was serendipitous, as the authors of \cite{ludwig:2015pl} were initially searching for an $f(R,\Lm)$ theory exhibiting \textit{intrinsic decoupling}, similar to what had been observed in certain Einstein-dilaton theories \cite{minazzoli:2013pr,minazzoli:2014pr,minazzoli:2014pl}. The rationale behind this search was based on the well-established fact that $f(R)$ theories are mathematically equivalent to specific Scalar-Tensor theories. This led the authors of \cite{ludwig:2015pl} to hypothesize that an $f(R,\Lm)$ theory might exist that would correspond to Einstein-dilaton theories with \textit{intrinsic decoupling}. However, the inherently non-linear nature of the resulting action was completely unexpected and was discovered almost by accident while manipulating the field equations.

A key conceptual advantage of Entangled Relativity over General Relativity is that it requires fewer fundamental constants at both classical and quantum levels \cite{minazzoli:2022ar}. However, an important consequence of this feature is that both $G$ and $\hbar$ are not fixed but can vary across space and time, leading to potential quantitative ways to test the theory \cite{minazzoli:2022ar,chehab:2025hl}.

The objective of this paper is to demonstrate that the non-linear action of Entangled Relativity could have been derived---rather than discovered accidentally---by imposing a straightforward requirement: the theory must possess all the solutions of General Relativity without a cosmological constant\footnote{For now on, when we mention ``General Relativity'', we mean ``General Relativity without a cosmological constant'', unless specified otherwise.} whenever $\Lm = T$ on-shell, similarly to the \textit{instrinsic decoupling} originally found in Scalar-Tensor theories \cite{minazzoli:2013pr,minazzoli:2014pr,minazzoli:2014pl}. 

The motivation for this requirement is that our present universe appears to be predominantly filled with dust and electromagnetic radiation, both of which satisfy $\Lm = T$ on-shell.\footnote{But see also \cite{avelino:2022pr,pinto:2025ar}.} Therefore, this condition is essential for recovering the well-tested phenomenology of General Relativity for our universe in his current state \cite{wambsganss:1998lr,uzan:2011lr,will:2014lr,abbott:2020lr,burns:2020lr,fienga:2024lr}, as has been confirmed within the framework of Entangled Relativity \cite{ludwig:2015pl,minazzoli:2018pr,minazzoli:2021cq,minazzoli:2024mo,arruga:2021pr,arruga:2021ep,minazzoli:2021ej,minazzoli:2022ar,minazzoli:2025ep,wavasseur:2025gg,minazzoli:2025ep}.

For definiteness, in what follows we refer to as an \textit{intrinsic decoupling} when the equation on the scalar degree of freedom $f_R:=\partial f/\partial R$ that comes from the non-linearity of the function $f(R,\Lm)$ is not sourced for matter fields that are such that $\Lm=T$ on-shell.

\section{Derivation}

Let us start from the most general $f(R,\Lm)$ theory, whose quantum phase reads as follows
\be
\Theta  =  \frac{1}{\xi_f} \int d^4_g x f(R,\Lm),
\ee
where one has defined the quantum parameter $\xi_f$ whose dimension is $[\xi_f] = [f] \times L^4$, such that $[\Theta]  = \emptyset$. $f$ is assumed to be a differentiable function of $R$ and $\Lm$. If $f$ is affine in $\Lm$, then $\xi_f$ has the dimension of an energy times a length and its value must be $\xi_f = \hbar c$ in order to recover standard quantum physics in the limit where the effects of gravity can be neglected---such as in particle physics.\\

Classical physics derives from paths with stationary phases, $\delta \Theta = 0$, such that the metric field equation notably reads as follows \cite{harko:2010ep,harko:2013pr}
\bea\label{eq:metricf}
f_R &&\left(R_{\mu \nu} - \frac{1}{2} g_{\mu \nu} R \right) + \left( g_{\mu \nu}\Box - \nabla_\mu \nabla_\nu \right)f_R \nonumber \\
&&-\frac{1}{2} g_{\mu \nu} \left(f-f_{\Lm} \Lm - f_R R \right) = \frac{f_{\Lm}}{2} T_{\mu \nu}.
\eea
Obviously, the quantum parameter $\xi_f$ has no impact on the classical field equations. In some sense, $\xi_f$ plays the role of the quantum of action $\hbar$ in standard physics.\\

Let us now show that demanding that the theory possesses all the solutions of General Relativity, for matter fields that are such that $\Lm=T$ on-shell, single out Entangled Relativity.\\

Given that the equation of General Relativity reads
\be
R_{\mu \nu} - \frac{1}{2} g_{\mu \nu} R = \kappa T_{\mu \nu},
\ee
where $\kappa$ is the coupling constant of General Relativity, if all solutions of General Relativity for matter fields that are such that $\Lm = T$ on-shell are also solutions of the $f$ theory, it means that one must have
\be\label{eq:fulleqtyb}
g_{\mu \nu} \left[\Box f_R -\frac{1}{2}  \left(f-f_{\Lm} \Lm - f_R R \right)  \right] =  \nabla_\mu \nabla_\nu f_R,
\ee
in Eq. (\ref{eq:metricf}), as well as
\be \label{eq:kappa}
\frac{f_{\Lm}}{2 f_R} = \kappa.
\ee

\subsection{Constant scalar degree of freedom: an \textit{intrinsic decoupling} is inescapable}

 Using its own trace, Eq. (\ref{eq:fulleqtyb}) can be rewritten as follows
\be \label{eq:fulleqty}
g_{\mu \nu} \left(f-f_{\Lm} \Lm - f_R R \right) = 6 \nabla_\mu \nabla_\nu f_R.
\ee
This equation demands that $\nabla_\mu \nabla_\nu f_R \propto g_{\mu \nu}$ for all the considered solutions and in all coordinate systems. This is only possible if $f_R$ is constant, as one can see directly in Riemann normal isotropic coordinates \cite{MTW}, where one would need to have $\partial_\mu \partial_\nu f_R \propto \eta_{\mu \nu}$ for all solutions, at any given point defining the Riemann normal isotropic coordinate system.\\

Since $f_R$ must be constant, it implies that Eq. (\ref{eq:fulleqty}) must reduce to
\be\label{eq:eqdemand1}
f-f_{\Lm} \Lm - f_R R = 0.
\ee
Thanks to Euler's homogeneous function theorem, we determine that the solution to this equation is a positively homogeneous function of degree 1. The general solution for this equation is
\bea \label{eq:sold1}
f(R,\Lm) = g \left(\frac{R}{\Lm} \right) \Lm + h\left(\frac{\Lm}{R} \right) R,
\eea
where $g(x)$ and $h(x)$ arbitrary dimensionful differentiable functions of $x$. Let us note that General Relativity obviously belongs to this class of theories. It notably corresponds to $g=1$ and $h=1/2\kappa$.\\

Now, one has to make sure that a constant $f_R$ can indeed be solution of the whole field equations for all matter fields that are such that $\Lm=T$ on-shell. In order to study this aspect, one simply has to consider the trace of Eq. (\ref{eq:metricf})---taking into account Eq. (\ref{eq:eqdemand1})---which provides the dynamical equation on the additional scalar degree of freedom that comes from the non-linearity of $f$:
\be
 3 \Box f_R = \frac{f_{\Lm}}{2} T+f_R R = 0.
\ee
For $\Lm=T$ on-shell, the decoupling condition $\Box f_R =0$ that allows for a constant scalar degree of freedom reduces to
\be \label{eq:fTrace}
\frac{f_{\Lm}}{2} \Lm+f_R R = 0.
\ee
Combining Eqs. (\ref{eq:eqdemand1}) and (\ref{eq:fTrace}), one obtains the following set of equations:
\bea
\begin{cases}
f+f_R R = 0, \\
2f - f_{\Lm} \Lm =0,
\end{cases}
\eea
whose general solution turns out to correspond to Entangled Relativity:
\be
f(R,\Lm) = C~ \frac{\Lm^2}{R},
\ee
with $C$ a constant.\\

\subsection{$\kappa$ is also constant}

It is important to check Eq. (\ref{eq:kappa})---that is, that  $f_{\Lm}$ is also constant in Eq. (\ref{eq:metricf}) when $f_R$ is constant, since $f_{\Lm}/(2f_R)$ has to correspond to the coupling constant $\kappa=8\pi G/c^4$ of General Relativity. One can check that it is indeed the case because $f_{\Lm} = 2 \sqrt{-f_R}$ for $f\propto \Lm^2/R$. In particular, it implies that the ratio between $R$ and $\Lm$ remains constant---even in the limit $\{R, \Lm\} \rightarrow 0$---which, in fact, should not be too surprising. Indeed, in General Relativity, the ratio between $R$ and $T$ is constant for all values of $R$ and $T$.\\

As a consequence, all the solutions of General Relativity with matter fields that are such that $\Lm=T$ on-shell are also solutions of Entangled Relativity, and it is the only $f(R,\Lm)$ theory that has this property.\\

Let us note, however, that even for matter fields satisfying $\Lm = T$, Entangled Relativity admits more solutions than General Relativity, since one can impose boundary or initial conditions with a non-zero derivative of the scalar degree of freedom. Moreover, for matter fields such that $\Lm \neq T$,\footnote{Such as an electric field for instance, where $\Lm \propto E^2$ while $T = 0$ \cite{minazzoli:2021ej,minazzoli:2025ep,minazzoli:2025ej}.} the solutions of Entangled Relativity generally differ from those of General Relativity. Therefore, in general, the theory remains distinct from General Relativity.

\section{Recovering General Relativity from Entangled Relativity}

The quantum phase that defines Entangled Relativity reads as follows \cite{minazzoli:2022ar};
\be
\Theta = -\frac{1}{2 \epsilon^2} \int d^4_g x \frac{\L^2_m}{R}, 
\ee
where $\epsilon$ is the reduced Planck energy \cite{minazzoli:2022ar,chehab:2025hl}. Classical physics follows from paths with stationary phases $\delta \Theta =0$. But it has been known since its discovery in \cite{ludwig:2015pl} that the whole set of equations that follow from $\delta \Theta =0$ can be recovered by $\delta \Theta_{Ed}=0$ instead, where $\Theta_{Ed}$ is an alternative Einstein-dilaton phase 
\be \label{eq:ER2QP}
\Theta_{Ed} = \frac{1}{\epsilon^2} \int d^4_g x \frac{1}{\kappa}\left(\frac{R}{2 \kappa} + \Lm \right),
\ee
provided that $\L_m \neq \emptyset$, and where $\kappa$ is a dimensionful scalar-field. In other words, $\Theta$ and $\Theta_{Ed}$ describe the same theory at the classical level.\\

Our present universe appears to be predominantly filled with dust. From the classical equivalence between $\Theta$ and $\Theta_{Ed}$, it is patent that Entangled Relativity behaves identically to General Relativity in a universe composed solely of a dust field, as in that case, one would have
\be
\Theta_{Ed} = \frac{1}{\epsilon^2} \int d^4_g x \frac{R}{2 \kappa^2} - \sum_i \int \frac{m_i c^2}{ \kappa \epsilon^2}~cd\tau,
\ee
where $m_i$ are the conserved mass of the particle $i$ along their trajectories ($dm_i/d\tau=0$) \cite{minazzoli:2013pd}.
Therefore, by performing the conformal transformation $\tilde g_{\alpha \beta} = (\bar \kappa / \kappa)^2 g_{\alpha \beta}$, where $\bar \kappa$ is simply a normalization constant, the phase reduces to
\bea
\Theta_{Ed} &=&\frac{1}{\bar \kappa \epsilon^2} \int  d^4x \sqrt{-\tilde g} \frac{1}{2 \bar \kappa}\left(\tilde R - 2 \tilde g^{\alpha \beta} \partial_\alpha \varphi \partial_\beta \varphi \right) \nonumber\\
&&- \sum_i \frac{m_i c^2}{ \bar \kappa \epsilon^2} \int ~cd \t \tau, \label{eq:actionEinstein}
\eea
where $\varphi$ is an uncoupled scalar field defined as $\varphi := \sqrt{3}~ \ln (\kappa / \bar \kappa) +$ constant \citep{arruga:2021ep,minazzoli:2021ej,wavasseur:2025gg,minazzoli:2025ep}, and $d \t \tau = (\bar \kappa / \kappa) ~d\tau$---such that $d m_i / d\t \tau =0$. Thus, in this scenario, Entangled Relativity is effectively indistinguishable from General Relativity, apart from an additional uncoupled—and therefore physically irrelevant—scalar field \citep{minazzoli:2015ar}.\\

This is due to the \textit{intrinsic decoupling} mentioned previously, which decouples the gravitational scalar degree of freedom whenever $\Lm=T$ on-shell. Indeed, one has 
\be
\Lm = - \sum_i \frac{m_i c^2}{\sqrt{-g} u^0} \delta^{(3)}(\vec{x} - \vec{x}_i(t)) = T,
\ee
for a dust field, with $u^\alpha = dx^\alpha/d\tau$.\\

Of course, our present universe is not composed entirely of dust, and the situation is more complex than this \cite{minazzoli:2018pr,minazzoli:2021cq,minazzoli:2024mo}. Nevertheless, this example provides a good intuition for why the phenomenology of Entangled Relativity should not significantly deviate from that of General Relativity at late times of our universe.

 \section{The vacuum limit}

A vacuum solution cannot be defined in Entangled Relativity.\footnote{Fortunately, there does not exist a single location in the universe where the value of the matter fields would be \textit{exactly} zero. In other words, \textit{a vacuum}, in the classical sense, does not exist in nature.} Indeed, in it, the scalar degree of freedom $f_R$ that comes from the non-linearity of $f$ is proportional to the square of the ratio between $\Lm$ and $R$, which is ill-defined for $R=\Lm=0$. Therefore, Entangled Relativity better reflects Einstein's ``Mach's Principle'' than General Relativity---which states that there can be \textit{no spacetime without matter} \cite{einstein:1917co,einstein:1918an,einstein:1918sp,einstein:1921bk,pais:1982bk,hoefer:1995cf,minazzoli:2024pn}.\\

Nevertheless, the ratio between $\Lm$ and $R$ is perfectly well defined and finite in the vacuum limit $\{R,\Lm,T_{\mu\nu}\} \rightarrow 0$, since $f_R$ obeys a scalar-field equation that reduces to an unsourced equation $\Box f_R \rightarrow 0$ in that limit. It has been verified with several exact solutions already \cite{minazzoli:2021ej,wavasseur:2025gg,minazzoli:2025ep}, and recently explained more generally in \cite{minazzoli:2025ej}.\\

As a consequence, all vacuum solutions of General Relativity are limits of solutions of Entangled Relativity when the matter fields tend to zero \cite{minazzoli:2025ej}.

 \section{An \textit{intrinsic decoupling} is not sufficient to recover General Relativity: on a general class of theories with \textit{intrinsic decoupling}}
\label{sec:genindec}

Because one needs to have a constant scalar degree of freedom $f_R$ to satisfy the demand that all the solutions of General Relativity are also solutions of the $f$ theory, an \textit{intrinsic decoupling} is necessary. Let us stress however that it is not sufficient.\\

Indeed, one could have started by demanding that Eq. (\ref{eq:metricf}) allows for a constant scalar degree of freedom whenever $\Lm=T$ on-shell. Again, the equation that drives this scalar degree of freedom comes from the trace of Eq. (\ref{eq:metricf}) and reads as follows
\be
3 \Box f_R = 2f - \frac{3}{2} f_{\Lm} \Lm - f_R R.
\ee
Hence, for that degree of freedom to be allowed to be constant, one needs to have
\be
2f - \frac{3}{2} f_{\Lm} \Lm - f_R R = 0.
\ee

Thanks again to Euler's homogeneous function theorem, we determine that the solution to this equation is a positively homogeneous function of degree 1. The general solution for this equation is
\be\label{eq:solCfR}
f(R,\Lm) = R^2~ q \left(\frac{R}{\Lm^{2/3}}\right),
\ee
where $q(X)$ is an arbitrary differentiable function of $X$. Entangled Relativity obviously belongs to this class of solutions. It corresponds to $q(X)\propto X^{-3}$.\\

The theories provided by Eq. (\ref{eq:solCfR}) all have solutions with a constant scalar degree of freedom $f_R$ whenever $\Lm=T$ on-shell, but only Entangled Relativity also possesses all the solutions of General Relativity. This is because 
\be\label{eq:neqfulleqtyb}
f-f_{\Lm} \Lm - f_R R \neq 0
\ee
in general for those theories. Indeed, considering the solutions for which $f_R$ is constant, Eq. (\ref{eq:metricf}) reduces to
\be \label{eq:aGRCC}
R_{\mu \nu} - \frac{1}{2} g_{\mu \nu} R  +\t \Lambda g_{\mu \nu} = \t \kappa T_{\mu \nu},
\ee
with 
\bea\label{eq:tlambda}
\t \Lambda &:=& -\frac{1}{2 f_R} \left(f-f_{\Lm} \Lm - f_R R \right), \nonumber\\
&=& \frac{R }{4}~\frac{q + \frac{1}{3} \frac{R}{\Lm^{2/3}} q'}{ q + \frac{1}{2} \frac{R}{\Lm^{2/3}} q'}
\eea
and
\be \label{eq:tkappa}
\t \kappa := \frac{f_{\Lm}}{2 f_R} = -\frac{1}{3} \frac{R^2}{\Lm^{5/3}} \frac{q'}{2q+\frac{R}{\Lm^{2/3}} q'}.
\ee
Let us stress that neither $\t \Lambda$ nor $\t \kappa$ are constant in general---despite $f_R$ being constant---such that the solutions of general relativity with a cosmological constant are not solutions of Eq. (\ref{eq:aGRCC}) in general. However, one verifies again that it is the case for Entangled Relativity, which corresponds to $q(X) \propto X^{-3}$, because $f_{\Lm}$ becomes a function of $f_R$ in that case, and $\t \Lambda =0$.\\

Let us note otherwise that Eq. (\ref{eq:solCfR}) is a generalisation of what was presented in \cite{minazzoli:2024mo}, where it was noticed that a theory of the following form
\be  \label{eq:EERp}
f(R,\Lm) = \sum_{n} \frac{1}{\epsilon^{2n}}\frac{\omega_n}{2n} \frac{\Lm^{2n}}{R^{3n-2}},
\ee
where $\omega_n$ are dimensionless constants, leads to an \textit{intrinsic decoupling} of the scalar degree of freedom---that is, $\Box f_R =0$ for $\Lm=T$---and therefore has solutions to an equation of the form of Eq. (\ref{eq:aGRCC}). Indeed, one can check that assuming $q(X) = \sum_n a_n (X^{-3})^n$ in Eq. (\ref{eq:solCfR}) leads to Eq. (\ref{eq:EERp}), with $a_n$ defined accordingly. Such series, if infinite, could correspond to the expansion of any analytical function, such as $q(X) = e^{a/(\epsilon^2 X^3)}$ for instance, where $a$ is an arbitrary dimensionless constant. Indeed, let us stress that $\epsilon^2 X^3 = \epsilon^2 R^3 / \Lm^2$ is adimensional.\\

One can check that beyond Entangled Relativity, a free Starobinsky term $f(R,\Lm) \propto R^2$ is another specific case of Eq. (\ref{eq:solCfR}) that allows for $\t \Lambda$ and $\t \kappa$ to be constant, as it implies $R = 4\t \Lambda$ and $\t \kappa=0$. Notably, all (anti-)de-Sitter solutions are solutions of this specific theory---since Eq. (\ref{eq:metricf}) reduces to $R_{\mu \nu}=g_{\mu \nu} R/4$ for any constant $R$ in that case.\\

More generally, theories with an \textit{intrinsic decoupling} are interesting because, as explained in \cite{minazzoli:2024mo}, the scalar degree of freedom automatically freezes for a universe in expansion that is filled with dust and electromagnetic radiation like our universe. Indeed, assuming a Friedmann-Lemaître-Robertson-Walker metric, the equation on the scalar degree of freedom in that situation reduces to
\be \label{eq:friction}
\ddot f_R + 3 H \dot f_R = 0 ,
\ee
where $H:=\dot a / a$ is the Hubble parameter. Therefore, the friction term from the cosmological expansion at the beginning of the matter dominated era freezes the scalar degree of freedom no matter what, such that one has
\be \label{eq:frozen}
\dot f_R \propto a^{-3}.
\ee
As a consequence, Eq. (\ref{eq:aGRCC}) is an inevitable limit of Eq. (\ref{eq:metricf}) for the theories defined from Eq. (\ref{eq:solCfR}), as soon as one considers a universe in expansion and filled with dust and electromagnetic radiation.\\

Nevertheless, it is hard to imagine that one of the theories described by Eq. (\ref{eq:solCfR}) other than Entangled Relativity could lead to a phenomenology consistent with observations, since $\t \Lambda$ and $\t \kappa$ do not appear to be allowed to be constant in general. A thorough investigation is left for future work.

\section{Conclusion}

Entangled Relativity is the only $f(R,\Lm)$ theory that admits all the solutions of General Relativity without a cosmological constant for matter fields satisfying $\Lm = T \neq 0$ on-shell---such as in a universe filled with dust and electromagnetic radiation, like ours. All other $f(R,\Lm)$ theories---including the subclass of the form $f(R,\Lm) = F(R) + \Lm$---typically require either fine-tuned parameters, dynamical decoupling mechanisms, or both \cite{uzan:2011lr,will:2014lr,burrage:2018lr,fienga:2024lr}, in order to reproduce the well-tested phenomenology of General Relativity.

The problem with such dynamical decoupling mechanisms---as opposed to an \textit{intrinsic decoupling}---is that they may succeed in certain dynamical regimes but fail in others. As a result, these models often struggle to account for all observational successes of General Relativity simultaneously, sometimes even after significant parameter tuning (see, e.g., \cite{hees:2012pr,burrage:2018lr}). In contrast, an \textit{intrinsic decoupling} works as long as the matter fields satisfy $\Lm = T$ on-shell.

Interestingly, Entangled Relativity is defined upon less fundamental constants than General Relativity \cite{minazzoli:2022ar,chehab:2025hl} and therefore is preferable from Occam's razor perspective. Given that it is more economical in terms of fundamental constants and units than General Relativity \cite{minazzoli:2022ar}, and considering that General Relativity has been confirmed with high precision across a wide range of phenomena \cite{wambsganss:1998lr,uzan:2011lr,will:2014lr,abbott:2020lr,burns:2020lr,fienga:2024lr}, we argue that Entangled Relativity should be regarded as a serious alternative to Einstein's general theory of relativity. 

The question of the acceleration of the universe's expansion remains open in Entangled Relativity \cite{minazzoli:2018pr,minazzoli:2021cq,minazzoli:2024mo}. However, it now appears to be an open question for General Relativity with a cosmological constant as well, given that the Hubble tension has reached significant levels across multiple lines of investigation \cite{verde:2024an}, among other anomalies \cite{peebles:2025pt}. However, unlike in General Relativity, the $\Lambda$CDM model cannot be a solution in Entangled Relativity \cite{minazzoli:2018pr,minazzoli:2021cq,minazzoli:2024mo}. Therefore, growing evidence against $\Lambda$CDM would place both theories on more equal footing with respect to this particular question. Another potential direction of research would be to explore the broader class of theories featuring an \textit{intrinsic decoupling}, as introduced in Sec. \ref{sec:genindec}.

Finally, Entangled Relativity predicts slight deviations from General Relativity at the densities of neutron stars and white dwarfs that could become measurable in the foreseeable future \cite{minazzoli:2022ar,chehab:2025hl}, notably from the observation of white dwarfs' spectra \cite{chehab:2025hl}, as already investigated in other contexts in \cite{bainbridge:2017un,hu:2019mn,lee:2022ar,lee:2024ax}. Perhaps even more interestingly, it has recently been shown that traversable wormhole solutions that satisfy the \textit{null energy conditions} exist in theories that can be written as an Einstein-Maxwell-dilaton theory \cite{bixano:2025pr,bixano:2025ar,bixano:2025ax,bixano:2025as}. Among such theories---such as 5D Kaluza-Klein theory---Entangled Relativity stands out as the only one that remains observationally viable \citep{ludwig:2015pl,minazzoli:2018pr,arruga:2021pr,arruga:2021ep,minazzoli:2021ej,minazzoli:2021cq,minazzoli:2024mo,wavasseur:2025gg,minazzoli:2025ep}, thanks to its \textit{intrinsic decoupling}. Therefore, Entangled Relativity opens up the possibility of the existence of wormholes in nature.


\bibliography{ER_derivation}


\end{document}